\documentclass[letterpaper, 10 pt, conference]{ieeeconf}  

\IEEEoverridecommandlockouts                              

\usepackage{cite}
\usepackage{amsmath,amssymb,amsfonts}
\usepackage{graphicx}
\usepackage{textcomp}
\usepackage{xcolor}
\usepackage{verbatim}
\usepackage{placeins}
\usepackage{dirtytalk}
\usepackage[ruled,vlined]{algorithm2e}
\usepackage{mathtools}
\usepackage{soul}
\usepackage{url}
\usepackage{graphicx}
\usepackage{subfig}
\usepackage{tikz}

\usepackage{xcolor}
\usepackage{algpseudocode}
\usepackage{footnote}

\makeatletter
\let\NAT@parse\undefined
\makeatother
\usepackage{hyperref}

\hypersetup{
  colorlinks,
  citecolor= red,
  linkcolor=blue,
  urlcolor=blue}

\title{\LARGE \bf
Black-Box Safety Validation of Autonomous Systems: A Multi-Fidelity Reinforcement Learning Approach
}

\author{Jared J. Beard$^{1}$ and Ali Baheri$^{2}$
\thanks{$^{1}$Jared J. Beard is a Ruby Distinguished Doctoral Fellow with the Mechanical and Aerospace Engineering Department at West Virginia University, Morgantown, WV 26505, USA.
        {\tt\small jbeard6@mix.wvu.edu}}%
\thanks{$^{2}$Ali Baheri is with the Mechanical Engineering Department at Rochester Institute of Technology, Rochester, NY 14623, USA.
        {\tt\small akbeme@rit.edu}}%
}

\begin{document}

\maketitle
\thispagestyle{empty}
\pagestyle{empty}

\begin{abstract}

The increasing use of autonomous and semi-autonomous agents in society has made it crucial to validate their safety. However, the complex scenarios in which they are used may make formal verification impossible. To address this challenge, simulation-based safety validation is employed to test the complex system. Recent approaches using reinforcement learning are prone to excessive exploitation of known failures and a lack of coverage in the space of failures. To address this limitation, a type of Markov decision process called the \say{knowledge MDP} has been defined. This approach takes into account both the learned model and its metadata, such as sample counts, in estimating the system's knowledge through the \say{knows what it knows} framework. A novel algorithm that extends bidirectional learning to multiple fidelities of simulators has been developed to solve the safety validation problem. The effectiveness of this approach is demonstrated through a case study in which an adversary is trained to intercept a test model in a grid-world environment. Monte Carlo trials compare the sample efficiency of the proposed algorithm to learning with a single-fidelity simulator and show the importance of incorporating knowledge about learned models into the decision-making process.

\end{abstract}

\section{INTRODUCTION}

The process of safety validation checks if a system meets performance standards or identifies the nature of potential failures. This is crucial in situations where human safety is at risk or significant harm to expensive equipment may occur, such as in aircraft autopilots, driverless cars, and space-flight systems. Formal verification, which is a commonly used method, builds a detailed mathematical or computational model of the system being tested to ensure it meets safety specifications. However, a major challenge in using formal methods for safety validation is the large and complex design space of decision-making models, which makes it difficult for formal verification techniques to work effectively.

\begin{figure}[t!]
\centerline{\includegraphics[width=\linewidth,height=\textheight,keepaspectratio]{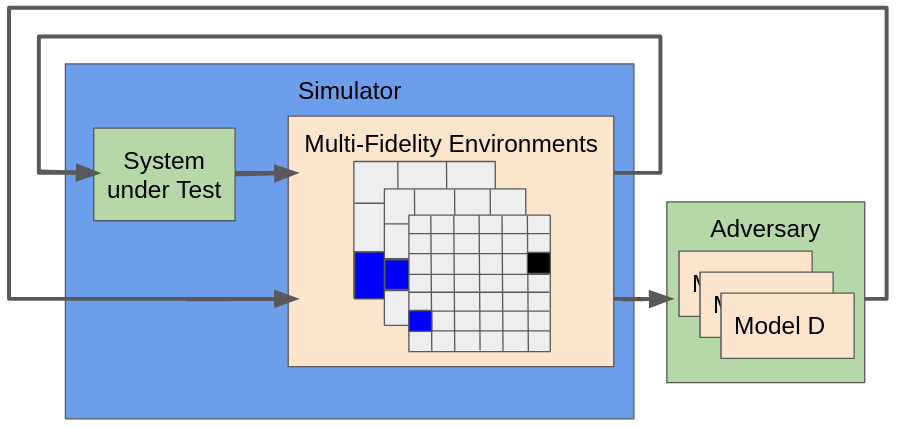}}
    \centering
    \caption{The \texttt{MF-RL-falsify} algorithm is applied to a grid world scenario where the aim is to train an adversary to intercept a system under test, which is defined as \emph{failure} in this study. The lowest fidelity having an unknown policy is selected and the corresponding simulator is sampled at each time step. Information about learned models is used to augment the reward structure, preventing early convergence of the algorithm and ensuring continued exploration for failures.}
    \label{fig:mf_kwik_ast}
\end{figure}

Another way of performing safety validation is through black-box and grey-box safety verification, where the system is either not known well enough or only provides limited information for analysis, respectively \cite{corso2021survey}. These methods can include planning algorithms \cite{plaku2009hybrid,tuncali2019rapidly}, optimization tools \cite{mullins2018adaptive,deshmukh2017testing}, and reinforcement learning \cite{lee2020adaptive}. From a reinforcement learning viewpoint, the goal of adaptive stress testing (AST) is to determine the most probable failure scenarios by treating the problem as a Markov decision process (MDP) and finding an adversarial policy that associates environmental states with disturbances that result in system failure. AST has been applied in various real-world domains, from aircraft collision avoidance \cite{lee2020adaptive} to autonomous vehicles \cite{koren2018adaptive}. Despite its overall success, sampling-based methods such as reinforcement learning have two main limitations. They may not be able to find rare failures \cite{corso2020scalable}, and they rely on data collected from high-fidelity simulators, which may not be feasible due to computational constraints \cite{moss2020adaptive}.

The requirement of relying solely on samples from high-fidelity simulators can be reduced by incorporating reinforcement learning techniques that utilize multiple levels of simulator fidelity \cite{cutler2015real}. The degree of accuracy of a simulator is referred to as its fidelity, with the level of fidelity representing how much the system model and assumptions are simplified. Low-fidelity simulators make strong assumptions, leading to faster execution, but may not exhibit realistic behavior and dynamics. On the other hand, high-fidelity simulators aim to closely approximate reality, but their execution may be slower due to more complex system models and fewer assumptions.

\noindent \textbf{Related work.} To date, there has been only one study on multi-fidelity (MF) safety validation, as described in \cite{koren2021finding}. This work serves as the only existing MF alternative to our proposed approach, where information is only passed from low to high-fidelity simulators in a unidirectional manner. However, unidirectional techniques limit the amount of information that can be passed, which affects their efficiency. On the other hand, bidirectional approaches like the one presented in this paper, which pass information in both directions, have been found to have lower sample complexity. The upper bound of high-fidelity samples in bidirectional approaches is equivalent to the unidirectional case \cite{cutler2015real}. Although this bidirectional approach is not used for safety validation, it provides valuable insights into the problem.

The bidirectional approach in this paper uses the \say{knows what it knows} (KWIK) framework to evaluate the relative worth of a sample in a given fidelity. The KWIK framework is a concept in reinforcement learning (RL) that involves explicitly tracking the uncertainty and reliability of learned models in the decision-making process. The RL algorithm in the KWIK framework is equipped with metadata about the learned models, such as sample counts, and uses this information to estimate the system's knowledge. This knowledge is then used to guide the exploration-exploitation trade-off in the RL algorithm.

\noindent \textbf{Statement of contributions.} The objective of this work is to develop a framework that minimizes the simulation cost while maximizing the number of failure scenarios discovered. In contrast to previous safety validation techniques, we propose a learner that uses knowledge about itself to reformulate the safety validation problem and present a bidirectional algorithm \texttt{MF-RL-falsify} to solve this problem (Fig. \ref{fig:mf_kwik_ast}). 
The contributions of this work are as follows:
\begin{itemize}
    \item We propose a novel knowledge Markov decision process (KMDP) framework, which factors knowledge (\textit{i.e.}, metadata) of the learned model into the decision making process and explicitly distinguishes the system model from the knowledge of this model;
    \item We demonstrate \texttt{MF-RL-falsify} an algorithm using the KWIK framework to train learners from \emph{multiple fidelities of simulators in a bidirectional manner}. \texttt{MF-RL-falsify} also uses a knowledge-based update to prevent early convergence of the learned model and encourage further exploration.
\end{itemize}

\noindent \textbf{Paper structure.}
The paper is structured as follows: Section \ref{sec:background} briefly outlines preliminaries, as well as introduces the KMDP. Section \ref{sec:approach} details the \texttt{MF-RL-falsify} algorithm. Section \ref{sec:results} covers the experimental setup, results, and some limitations of the implementation. Lastly, Section \ref{sec:conclusion} presents concluding remarks and future directions of work.


\section{Problem setup and Preliminary}\label{sec:background}

The objective of this work is to find the most likely failure scenarios for a (decision-making/control) system under test $\mathcal{M}$. Here, failure scenarios are defined as the sequence of transitions that lead to some failure condition (\textit{e.g.}, midair collision). The system under test is placed in a black-box (usually high-fidelity) simulator $\Sigma$, so the learner (with planner $P$) does not have direct access to the internal workings of the system. The learner then samples scenarios iteratively to learn and uses the planner to apply disturbances (\textit{e.g.}, control agents in the environment, supply noise, or adjust parameters of the environment). More specifically, the learner uses the simulations to minimize the distance to a failure metric. 
The work presented here seeks to reduce the need for computationally expensive simulations by using MF simulations with bidirectional information passing to improve the overall sample efficiency. 

\subsection{Markov Decision Process}
This work relies on MDP's so it is useful to introduce notation which will form the foundation of this work. MDP's are defined by the tuple $\langle\mathcal{S}, \mathcal{A}, T, R, \gamma \rangle$. Here, $\mathcal{S}$ is the set of all states $s$ and $\mathcal{A}$ the set of all actions $a$. The stochastic transition model is $T(s'\mid s,a)$, where $s'$ represents a state outcome when taking $a$ from $s$. $R(s,a,s')$ is the reward for some transition. Lastly, $\gamma$ is a temporal discount factor. By solving an MDP, an optimal control policy $\pi^*(s) \ \forall s \in \mathcal{S}$ can be achieved. Here, the optimal policy is defined as follows:
\begin{equation}
    \pi^*(s) = \arg\max_{a} Q(s,a),
\end{equation}
 where $Q^*(s,a)$ the corresponding optimal state-action value function is 
 \begin{multline}
    Q^*(s,a) =
    \\\max \sum_{s'}T(s'\mid s,a)[R(s,a,s') + \gamma Q^*(s',a')].
\end{multline}


\subsection{Knows What It Knows Learner} \label{sec:kwik}

Use of MF simulators in reinforcement learning has come about from the desire to reduce the number of samples in and expense incurred by high-fidelity simulators used to get reliable performance. To this end, \cite{cutler2015real} leveraged the concept of KWIK learners \cite{li2011knows}. The fundamental idea of KWIK is to use information about the learned model to provide output indicating its confidence in said model. When the learner is not confident in the model, it reports relevant state-action pairs as unknown ($\perp$) \cite{li2011knows}. In particular, Cutler, \textit{et al.} uses the number of samples to assert a model is known to some degree of confidence specified by the Hoeffding inequality (parameterized by $\epsilon$, $\delta$) \cite{cutler2015real}.

In the multi-fidelity case, the knowledge is used to decide when to \emph{switch} between the $D$ fidelities of simulators. The knowledge $\mathcal{K}$ is defined for this work as both estimates of the transition and reward models ($\mathcal{K}.\hat{T}$, $\mathcal{K}.\hat{R}$), as well as the confidence in these models \big($\mathcal{K}_{d,T}(s,a)$, $\mathcal{K}_{d,R}(s,a)$\big). Here, the subscript $d$ indicates the specific fidelity. The lower fidelities aim to guide search in higher fidelity simulators, while knowledge from higher fidelities is used to improve the accuracy of lower fidelity models. Mathematically, two simulators are said to be in fidelity to each other if their $Q$ values are sufficiently close according to: 
\begin{multline} 
    f(\Sigma_i, \Sigma_j, \rho_i, \beta_i) = \\
    \begin{cases}
          \Delta = - \max_{s,a} | Q^*_i(s,a) - Q^*_j(\rho_j(s),a)| , \\
          \hfill \forall s,a ~ \text{if} ~ \Delta \leq \beta_i \\
          -\infty, \  \mathrm{else.}
     \end{cases}
\end{multline}
Note, $\beta$ indicates a user specified bound on how dissimilar two simulators can be and $\rho$ is a mapping of states from the fidelity of $i$ to fidelity $j$. In this work, this mapping is applied on a state-by-state basis, as the decision process only needs to model one decision at a time, not the whole system.

\subsection{Knowledge Markov Decision Process}

Whereas this learning problem has classically been represented as an MDP, doing so assumes the model is known with sufficient accuracy. This may not always be the case (as in reinforcement learning). When the decision maker can quantify its lack of information about the model, it should be able to use this information to its benefit. Thus, we introduce the KMDP as $\langle\mathcal{S}, \mathcal{A}, T, R, \mathcal{K}, \gamma \rangle$. Here, $\mathcal{K}$ represents the knowledge about the decision maker's model of the system (\textit{e.g.}, estimates of the other terms $\langle\mathcal{S}, \mathcal{A}, T, R\rangle$, confidence in those estimates, and other assumptions of the learned model). Consequently, $\hat{T}$ and $\hat{R}$ are used to solve the underlying MDP as the agent learns. Note from Sec. \ref{sec:kwik} these estimates of the model are considered part of $\mathcal{K}$. In doing so, the learner is explicitly aware that it is making decisions based on an estimated model of its environment and that it has some domain knowledge regarding the quality of these estimates. Thus this information can factor into decisions, such as through the reward which is now defined as $\hat{R}(s,a,s',\mathcal{K})$. As an example, knowledge about the lack of information for some subset of states could necessitate a reward that encourages exploration in those regions. 

The KMDP is related to the concepts of maximum likelihood model MDPs and reward shaping. The maximum likelihood model MDP and KMDP share similar update rules for the learned reward and transition models \cite{wiering1998learning}. However, KMDP generalizes these concepts to include estimates of $\mathcal{S}$ and $\mathcal{A}$, as well as providing a more broad use of the information gained from sampling. As an example, the agent may not be aware of all states if it is required to explore its environment, thus the estimated state space would be a subset of $\mathcal{S}$. Furthermore knowledge may now capture concepts such as whether a state has been visited, which is not inherently captured by the underlying state. It is important to note that this would break the Markov property, but such an approximation has been used to break explicit temporal dependencies \cite{beard2020environment}. The second concept is reward shaping, where the additional reward terms guide learning as a function of the system state \cite{laud2004theory}. The difference being that here the learner is deciding based on learned information or knowledge about the model, not only the underlying state. Following from this, $Q^*$ cannot actually be achieved, and the best $Q$ given $\mathcal{K}$ becomes
 \begin{multline}
    Q_{\mathcal{K}}(s,a) = \max_a \sum_{s'} \\ \mathcal{K}.\hat{T}(s'\mid s,a)[\mathcal{K}.\hat{R}(s,a,s',\mathcal{K}) + \gamma Q_{\mathcal{K}}(s',a')].
\end{multline}

\section{Falsification using multi-fidelity simulators}\label{sec:approach}

Here, the functionality of the \texttt{MF-RL-falsify} algorithm (Alg. \ref{alg:mf_kwik_ast}) is described. The approach centers on three functionalities: \textit{search}, \textit{evaluateState}, and \textit{marginalUpdate}. The \textit{search} function is tasked with iterating through trials and collecting failure information. As part of \textit{search}, \textit{evaluateState} simulates trajectories and updates the learned models. Lastly, \textit{marginalUpdate}, given confidence information, updates the reward function to bias search away from trajectories for which the models are known (sampled a sufficient number of times). The proposed algorithm is initialized with a set of simulators, fidelity constants, and state mappings $\langle \Sigma, \beta, \rho \rangle$, the system under test $\mathcal{M}$, and the planner $P$ of choice (this work uses value iteration). Additionally, the confidence parameters ($\epsilon, \delta$), and minimum samples to change fidelity ($m_{known}$, $m_{unknown}$) are supplied. For more information on the choice of these parameters, see \cite{cutler2015real}. With no samples, $\mathcal{K}$ is initialized with a uniform distribution over transitions and rewards set as $R_{max}$ of the KMDP. The $Q$ values are set to 0 and the current fidelity $d$ is declared as the lowest fidelity. Lastly, it is assumed there are no changes to the model estimate, and no known or unknown samples $(m_k,m_u)$.

\begin{algorithm}
\caption{\texttt{MF-RL-falsify}}\label{alg:mf_kwik_ast}
\SetKw{Input}{input:}
\SetKw{Initialize}{initialize:}
\SetKwProg{Proc}{Procedure}{}{}

\Input{$\langle \Sigma, \beta, \rho \rangle$, $\mathcal{M}$, $R_{inc}$, $P$, $( \epsilon, \delta)$, $(m_{known}, m_{unknown})$}

\Initialize{ $\mathcal{K}$, $Q_d$}

\Initialize{ $change_d = false$, $d = 1$, $m_k,m_u = 0$}

\Proc{search($s$,$n$)}
{
$\mathcal{\hat{F}} \leftarrow \{\}$

\For {$i = 1, ..., n$}
{
    $reinitialize(\Sigma)$
    
    $\mathcal{\hat{F}} \leftarrow \mathcal{\hat{F}} \cup evaluateState(s)$    
}

\For {$f \in \mathcal{\hat{F}}$}
{
    \If {$\neg isPlausible(f,D)$}
    {
        $\mathcal{\hat{F}} \leftarrow \mathcal{\hat{F}} \setminus f$
    }
}
    
\Return $\mathcal{\hat{F}}$
}

\Proc{evaluateState(s)}
{
\Initialize{$f \leftarrow \{\}$}

\While{$\neg terminal(s)$}
{
    $a \leftarrow \{ P.\pi(s), \mathcal{M}.\pi(s)\}$
    
    \If{$d>1$ and $change_d$ and $\mathcal{K}_{d-1}(\rho_{d-1}(s),a_{P}) = \perp$ and $m_u \geq m_{unknown}$}
    {
        $Q_{d-1} \leftarrow plan(d-1)$

        $ d \leftarrow d-1$
    }
    \Else
    {
    
    $s',r \sim T_\Sigma(s'\mid s,a)$

    \If  {$\mathcal{K}_d(s,a) =$ $\perp $}
    {
        $update\big(\mathcal{K}_d(s,a)\big)$
       
    \If {$\mathcal{K}_d(s,a) \not =$ $\perp$ }
    {   
        $Q_{d} \leftarrow plan(d)$
            
    }
    }
        
    $f \leftarrow f \cup \langle s, a, s' \rangle$, $s \leftarrow s'$
    }
    
    \If{$d < D$ and $m_k \geq m_{known}$}
    {
        $Q_{d+1} \leftarrow plan(d+1)$
        
        $d \leftarrow d + 1$
    }
}
\If {isConverged($f)$}
{
    $marginalUpdate(f)$
}
\If {$isFailure(f)$}
{
\Return $f$
}

}
\Proc{plan(d)}
{
\For{ all $\langle s, a \rangle$}
{
    $d' \leftarrow \{max(d')\, \mid \, d' \geq d,\, \mathcal{K}_{d'}(s,a) \not = \perp\}$
    
    $\mathcal{K}_d(s,a) \leftarrow \mathcal{K}_{d'}(s,a)$ 
}

\Return $P.train(\mathcal{K}.\hat{T},\mathcal{K}.\hat{R},Q_d, Q_{d-1}+\beta_{d-1})$
}

\Proc{marginalUpdate($f$)}
{
$m_{max} \leftarrow 0,\, t \leftarrow \langle \rangle$

\For {$\langle s, a, s' \rangle \in f$}
{
    $m = Q_{\mathcal{K}}(s,a) - Q_{\mathcal{K}'}(s,a')$
    
    \If{$m > m_{max}$}
    {
        $m_{max} \leftarrow m$
        
        $t \leftarrow \langle s, a, s' \rangle$
    }
}

$R(t) \leftarrow R(t) - R_{inc}$

$Q_{d} \leftarrow plan(d)$

}
\end{algorithm}

\subsection{Search}

The $search$ algorithm serves primarily to initiate the learning process. \texttt{MF-RL-falsify} looks at failures from a single initial condition. As such, a state $s$ is passed in along with a desired number of trajectories $n$. The estimate of the set of failure modes, $\mathcal{\hat{F}}$ is initialized as empty. For every iteration, the simulators are reset and $evaluateState$ is called. Failure scenarios found by $evaluateState$ are added to $\mathcal{\hat{F}}$. Due to the nature of sampling, policies may come from more than one fidelity. Thus, after the search has concluded, it is necessary to evaluate whether failure modes are viable in the highest fidelity; those not meeting this condition are rejected. The remaining set is returned to the user. Plausibility checks may manifest in a few ways: for simulators with relatively simple transition models, these can be checked directly to exist or not; for more complex, stochastic systems, this may involve a Monte Carlo sampling to determine if a transition occurs in the model. Fortunately, due to the Markov assumption, this can be broken down to a state-by-state basis, as opposed to looking at the entire trajectory at once. 


\subsection{State Evaluation}

At every iteration of the $search$ algorithm, the \textit{evaluateState} function attempts to find a failure scenario given the current model estimate. The process begins by initializing the trajectory $f$. Then actions are iteratively selected from the current policy $\pi$ and the decision is made to increment the fidelity or sample the current one. This continues until a terminal state is found; terminal states can be a failure event, exceeding the maximum number of time steps, or reaching some state where failure cannot occur. Whenever the fidelity is incremented, both $m_k$ and $m_u$ are reset. Otherwise, if a known state-action is reached, $m_k$ is incremented, while $m_u$ is reset. The converse occurs when an unknown state is reached. Similarly, $change_d$ is set to false when the state is incremented and set to true when a state-action becomes known.

The fidelity is decremented if a known state has not recently been reached and the lower fidelity states are also unknown. This permits more efficient search by using the less expensive simulator to carry out uncertain actions. Along with the decrement, the planner is updated using the current model. Otherwise, a simulation is performed to gather data from the model. If the current state is considered unknown, this information is used to update the learned model and knowledge. The knowledge about the transition is incremented by one (the distribution is generated by normalizing the sampled outcomes for each $s,\,a$ pair). The reward is updated as a weighted average of the reward estimate and sampled reward. Lastly, the knowledge about each, being the number of samples for a state-action, is incremented by one. If a state-action becomes known, the planner is updated. The transition is added to the trajectory and the current state is updated. Subsequently, the fidelity is incremented and re-planning occurs if enough known samples have been visited. The algorithm concludes with the marginal update if the trajectory has converged \big(all $(s,a) \in f$ considered known\big). If a failure state has been achieved by the trajectory, the trajectory is deemed a failure scenario and returned. 

When planning, the algorithm must pass information to lower fidelities to improve the performance of future samples. The $plan$ function does this by finding the highest fidelity ($\geq d$) for which a state-action pair is known. This model estimate is then used to update the knowledge of the current fidelity. Similarly, to pass information up, if states from the current $d$ and next lower $d-1$ model estimates are deemed to be in fidelity ($\Delta \leq \beta$), the lower fidelity model estimate is used in place of the current fidelity. The planner is then run to update $Q$.

\subsection{Marginal Update}

The marginal update incorporates knowledge into the decision making process. Here, the marginal is defined as the difference in the best state-action value $Q_{\mathcal{K}}$ and the second best state-action value $Q_{\mathcal{K}'}$ for a given state. When the system is known to be converged, the state having the largest marginal (along the trajectory) is selected. Using this state, the reward is decreased by $R_{inc}$ and the planner updated. The underlying idea is that if the learner has converged on a trajectory, it will cease to explore the solution space, exploiting an action repeatedly. By finding the state where the best action is significantly greater than its neighbors, exploration can erode the most functionally important actions for a given trajectory. As training continues, this cuts off search in more fragile trajectories (hence less likely trajectories early) and encourages exploration along the entirety of more robust failure trajectories. 


\section{RESULTS}\label{sec:results}

We evaluate the effectiveness of the proposed algorithm by answering the following research questions:

\begin{itemize}

    \item \textbf{RQ1:} Does the proposed multi-fidelity algorithm improve the overall sample efficiency?\footnote{Sample efficiency refers to the ability of the RL agent to learn effectively from a relatively small number of environment interactions.}

    \item \textbf{RQ2:} Does the proposed multi-fidelity algorithm result in the identification of more failure scenarios compared to the single-fidelity scenario?
    
\end{itemize}
Before presenting the simulation setup, we define two key terms in our experiments.

\noindent{\textbf{Definition 1.} (Fidelity).} In this context fidelity refers to the accuracy of the model being used, with a high-fidelity model having a higher level of accuracy, capturing more details and nuances, while a low-fidelity model may not capture all the relevant information and result in less accuracy. In our experiments, the low-fidelity model does not accurately capture the impact of puddles on the transition model (see Fig. \ref{fig:env}).

\noindent{\textbf{Definition 2.} (Failure).} Here, failure is defined as the scenario where the red dot (MF-RL learner) intersects the blue dot (system under test).

\subsection{Simulation}

\begin{figure}[t!]
\centerline{\includegraphics[width=\linewidth,height=\textheight,keepaspectratio]{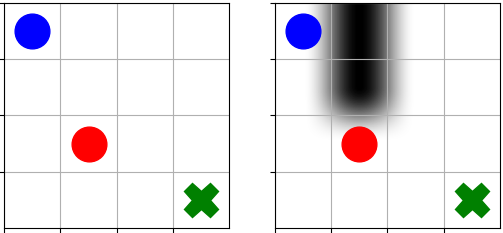}}
    \centering
    \caption{4$\times$4 Grid worlds under test (left) low-fidelity, (right) high-fidelity. Here, fidelity is demonstrated by accuracy of the model, where the low-fidelity does not capture the impact of puddles on the transition model. The blue dot indicates the system under test, the red the MF-RL learner, and the green ``$\times$'' the system under test's goal. Puddles are indicated by black and grey regions.}
    \label{fig:env}
\end{figure}

 The experiment involved a 4$\times$4 grid world environment with puddles (as depicted in Fig. \ref{fig:env}). The goal of the system under test was to reach a specific location in the grid world, while the learner's objective was to cause failures by either blocking the system under test or causing a collision. The system under test was designed to be myopic, meaning it would take the closest unobstructed action towards its goal. The states were defined by the coordinates of both the system under test $(x,y)_\mathcal{M}$ and the learner $(x,y)_P$. Both agents were able to move one cell in any of the cardinal directions or remain in place. If an agent was in a puddle, it had a 20\% chance of reaching the desired cell; otherwise, it would not move. The learner was penalized based on its distance from the system under test, receiving a penalty of $-5$ when in a puddle and $-25$ when the system under test reached the goal. If the learner induced a failure, it received a reward of $50$. The discount factor $\gamma$ was set to $0.95$ for the experiment. 

\subsection{Monte Carlo Trials}

To evaluate the approach, $25$ Monte Carlo trials were conducted for both the  single-fidelity (SF) and MF learners. 
To evaluate the performance of the marginal update, each of five values for $R_{inc} = [0, 0.25, 1, 2, 5]$ at up to $1000$ iterations were used in both simulators. Note that $R_{inc} = 0$ indicates no marginal update. Each trial, the initial coordinates of the system under test and learner were uniformly sampled; samples that obviously could not cause failures or scenarios initialized to failure states were removed. Other parameters were set as follows: $\beta = 1250$, $t_{max} = 20$, $\epsilon = 0.25$, $\delta = 0.5$, $m_{known} = 10$, and $m_{unknown} = 5$. 

\begin{figure}[t!]
\centerline{\includegraphics[width=\linewidth,height=\textheight,keepaspectratio]{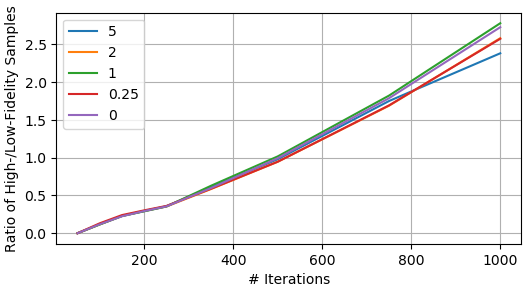}}
    \centering
    \caption{Mean ratio of high-fidelity samples to low-fidelity samples against iterations of $search$; notice that $R_{inc} = 5$ encourages more exploration as samples increase. Legend indicates values of $R_{inc}$.}
    \label{fig:sample_ratio}
\end{figure}

\begin{figure}[t!]
\centerline{\includegraphics[width=\linewidth,height=\textheight,keepaspectratio]{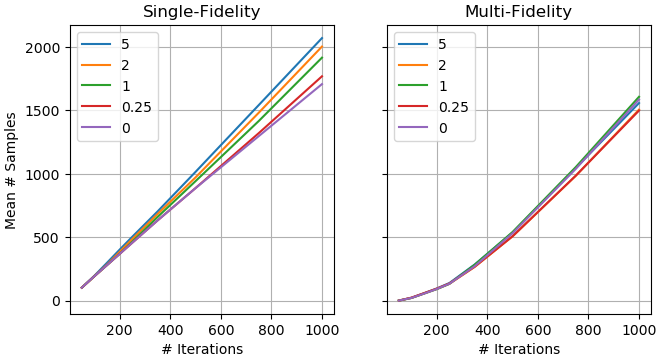}}
    \centering
    \caption{Mean number of high-fidelity samples for (left) single-fidelity and (right) Multi-fidelity learners. Notice the MF approach accumulates significantly fewer samples until convergence, at which samples are accumulated at similar rates. Legend indicates values of $R_{inc}$.}
    \label{fig:samples}
\end{figure}

Given the relatively simple scenario, it is helpful to understand where the MF simulator converges in the highest fidelity. Much larger examples involve sparse scenarios, so the area of interest are those regions where the ratio of high-fidelity to low-fidelity samples (Fig. \ref{fig:sample_ratio}) is relatively low and likely far below this convergence estimate. For context, the optimal high-fidelity policy could be found at a high- to low-fidelity sample ratio of less than $\sim$ 0.4. Notice that the marginal update does not appreciably impact the ratio, except for $R_{inc} = 5$, where the search reached under-explored states in low fidelities. Naturally, as the algorithm converges, return to lower fidelities is prevented and the samples grow at approximately the same rate as the SF case, hence the near linear behavior after convergence (Fig. \ref{fig:samples}) (\textbf{RQ1)}. In the SF scenario, increasing $R_{inc}$ led to further search, though in the MF scenario the behavior was unclear, likely due to interactions from changing fidelities.

\begin{figure}[t!]
\centerline{\includegraphics[width=\linewidth,height=\textheight,keepaspectratio]{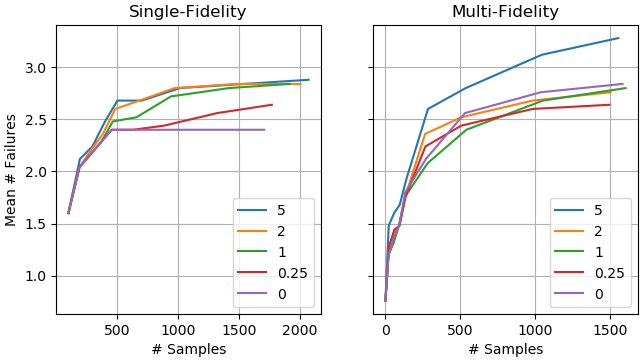}}
    \centering
    \caption{Mean number of high-fidelity failures for the (left) single- and (right) multi-fidelity approaches. The MF approach immediately makes use of underlying information, while the SF approach wastes early samples in the high-fidelity simulator. Additionally, $R_{inc}=5$ increases the number of failures found relative to no or a small update. Legend indicates values of $R_{inc}$.}
    \label{fig:failures}
\end{figure}

\begin{figure}[t!]
\centerline{\includegraphics[width=\linewidth,height=\textheight,keepaspectratio]{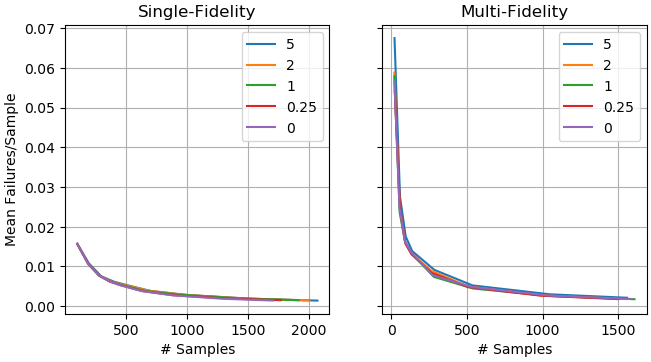}}
    \centering
    \caption{Ratio of failures per sample for the (left) single- and (right) multi-fidelity cases. The MF approach is approximately 6 times more efficient for few samples; their efficiency converges as samples increases. Additionally $R_{inc} = 5$ appears to provide an upper bound on the performance. Legend indicates values of $R_{inc}$.}
    \label{fig:failure_sample}
\end{figure}

With respect to the number of failures, the MF simulator quickly accumulates failures from low-fidelity samples, while the earliest high-fidelity samples in the SF simulator are wasted (Fig. \ref{fig:failures}) (\textbf{RQ2}). Around convergence, their performance is comparable. Without the marginal update, search in SF ceases at convergence; a large $R_{inc}$ yields $\sim 15\%$ increase in failures found. In the MF scenario, small updates perform similarly to no update, however, $R_{inc} = 5$ yields a significant improvement, as much as $15\%$ with almost no HF samples, supporting use of the marginal update. The improvements of the MF scenario are further realized in Fig. \ref{fig:failure_sample}. The MF scenario finds as many as 6 times more failures per sample than the SF case early on, with their performance approaching unity after convergence. 

Overall, results are promising, indicating the MF algorithm significantly improves the sample efficiency in finding failure scenarios over the SF approach. Additionally, the marginal update for large values of $R_{inc}$ enhances exploration consistently with growth in samples, leading to more failure scenarios and greater sample efficiency, particularly when the model has been sampled insufficiently (as in more complicated scenarios). 

\subsection{Limitations}

The implementation has two main drawbacks. One is that it is only created for discrete problems and handling representations of knowledge and continuous domains can be challenging within the KWIK framework. The second limitation is that value iteration, which is used to find solutions to MDPs, is not very efficient and thus limits the number of Monte Carlo trials and the size of the environment. This issue could potentially be resolved by using an anytime solver such as Monte Carlo tree search \cite{browne2012survey}.


\section{CONCLUSIONS} \label{sec:conclusion}

The purpose of this study was to enhance the effectiveness of reinforcement learning techniques in detecting failure scenarios in autonomous systems. The study introduced the knowledge MDP, which incorporates a learner's understanding of its own model estimates into the decision-making process, through the KWIK framework. The resulting algorithm, called \texttt{MF-RL-falsify} showed improved performance compared to single-fidelity (SF) approaches. The inclusion of knowledge through marginal updates was also found to increase exploration in converged scenarios and enhance early exploration of the solution space.

However, further development is needed for the technique to become a viable alternative to existing methods. This includes increasing scalability by utilizing learned models with anytime planners and applying the framework to continuous state and action spaces. Additionally, further research is necessary to test the KMDP formulation in various contexts and more complex scenarios.




\addtolength{\textheight}{-12cm}   






\bibliographystyle{IEEEtran} 
\bibliography{acc-mfrl}

\newpage

\section{APPENDIX} \label{sec:app}

The KWIK formulation (Alg. \ref{alg:kwik_ast}) was generated by stripping \texttt{MF-RL-falsify} of its dependence on fidelity. As such, supplying \texttt{MF-RL-falsify} with a SF simulator, will have the same performance. In spite of this, the KWIK algorithm was used for experiments. Notice, there is no longer a need to check the veracity of failures since all samples are from a single fidelity. Furthermore, planning is accomplished using the models for $\hat{T}$ and $\hat{R}$ as there is no longer information to pass. This reduces the $plan$ step to training as would be done in classical approaches.

\begin{algorithm}[bh!]
\caption{KWIK}\label{alg:kwik_ast}
\SetKw{Input}{input:}
\SetKw{Initialize}{initialize:}
\SetKwProg{Proc}{Procedure}{}{}

\Input{$\Sigma$, $\mathcal{M}$, $R_{inc}$, $P$, $( \epsilon, \delta)$}

\Initialize{ $\mathcal{K}$, $Q$}

\Proc{search($s$,$n$)}
{
$\mathcal{\hat{F}} \leftarrow \{\}$

\For {$i = 1, ..., n$}
{
    $reinitialize(\Sigma)$
    
    $\mathcal{\hat{F}} \leftarrow \mathcal{\hat{F}} \cup evaluateState(s)$    
}
    
\Return $\mathcal{\hat{F}}$
}

\Proc{evaluateState(s)}
{
\Initialize{$f \leftarrow \{\}$}

\While{$\neg terminal(s)$}
{
    $a \leftarrow \{ P.\pi(s), \mathcal{M}.\pi(s)\}$
    
    $s',r \sim T_\Sigma(s'\mid s,a)$
    
    \If  {$\mathcal{K}(s,a) =$ $\perp $}
    {
        $update\big(\mathcal{K}(s,a)\big)$
       
    \If {$\mathcal{K}(s,a) \not =$ $\perp$ }
    {   
        $Q \leftarrow P.train(\mathcal{K}.\hat{T},\mathcal{K}.\hat{R},Q)$
            
    }
    }
        
    $f \leftarrow f \cup \langle s, a, s' \rangle$, $s \leftarrow s'$
    }
    \If {isConverged($f)$}
    {
        $marginalUpdate(f)$
    }
    \If {$isFailure(f)$}
    {
        \Return $f$
    }

}

\Proc{marginalUpdate($f$)}
{
$m_{max} \leftarrow 0,\, t \leftarrow \langle \rangle$

\For {$\langle s, a, s' \rangle \in f$}
{
    $m = Q_{\mathcal{K}}(s,a) - Q_{\mathcal{K}'}(s,a')$
    
    \If{$m > m_{max}$}
    {
        $m_{max} \leftarrow m$
        
        $t \leftarrow \langle s, a, s' \rangle$
    }
}

$R(t) \leftarrow R(t) - R_{inc}$

$Q \leftarrow P.train(\mathcal{K}.\hat{T},\mathcal{K}.\hat{R},Q)$

}
\end{algorithm}

\end{document}